\documentclass[pra,twocolumn,aps,groupedaddress,showpacs]{revtex4}
\usepackage{graphicx}
\begin{document}

% Use the \preprint command to place your local institutional report
% number in the upper righthand corner of the title page in preprint mode.
% Multiple \preprint commands are allowed.
% Use the 'preprintnumbers' class option to override journal defaults
% to display numbers if necessary
%\preprint{}

%Title of paper
\title{Knight Field Enabled Nuclear Spin Polarization in Single Quantum Dots }

% repeat the \author .. \affiliation  etc. as needed
% \email, \thanks, \homepage, \altaffiliation all apply to the current
% author. Explanatory text should go in the []'s, actual e-mail
% address or url should go in the {}'s for \email and \homepage.
% Please use the appropriate macro for each type of information

% \affiliation command applies to all authors since the last
% \affiliation command. The \affiliation command should follow the
% other information
% \affiliation can be followed by \email, \homepage, \thanks as well.
\author{C. W. Lai}
\author{P. Maletinsky}
\author{A. Badolato}
\author{A. Imamoglu}
%\homepage[]{Your web page}
%\thanks{}
\affiliation{Institute of Quantum Electronics, ETH-Z\"{u}rich,
CH-8093, Z\"{u}rich, Switzerland}

\date{\today}

\begin{abstract}
We demonstrate dynamical nuclear spin polarization in the absence
of an external magnetic field, by resonant circularly polarized
optical excitation of a single electron or hole charged quantum
dot. Optical pumping of the electron spin induces an effective
inhomogeneous magnetic (Knight) field that determines the
direction along which nuclear spins could polarize and enables
nuclear-spin cooling by suppressing depolarization induced by
nuclear dipole-dipole interactions. Our observations suggest a new
mechanism for spin-polarization where spin exchange with an
electron reservoir plays a crucial role. These experiments
constitute a first step towards quantum measurement of the
Overhauser field.

\end{abstract}

% insert suggested PACS numbers in braces on next line

%\maketitle must follow title, authors, abstract, \pacs, and \keywords
\maketitle

% body of paper here - Use proper section commands
% References should be done using the \cite, \ref, and \label commands

Hyperfine interactions in quantum dots (QD) are qualitatively
different than those in atoms: coupling of a single electron-spin
to the otherwise well-isolated quantum system of nuclear spins in
a QD gives rise to rich physical phenomena such as non-Markovian
electron-spin decoherence
\cite{Khaetskii2002,Johnson2005,Koppens2005}. It has also been
proposed that the long-lived collective nuclear-spin excitations
generated and probed by hyperfine interactions could have
potential applications in quantum information processing
\cite{Taylor2003a,Taylor2003b}. Several groups have previously
reported QD nuclear-spin cooling using external magnetic fields
\cite{Gammon1997,Ono2004,Eble2005}. To achieve dynamical nuclear
spin polarization (DNSP), it has generally been assumed that a
small but nonzero external magnetic field is necessary.

Here, we use resonant circularly polarized optical excitation of a
single electron or hole charged QD to demonstrate DNSP in the
absence of an external magnetic field. We show that optical
pumping of the electron spin induces an effective inhomogeneous
magnetic (Knight) field that can be more than an order of
magnitude larger than the characteristic nuclear dipolar fields,
which in turn ensures that DNSP is not suppressed by the latter.
Our experiments constitute a first
step towards projective quantum measurements of the effective
nuclear (Overhauser) field operator that could in turn suppress electron
spin decoherence in QDs\cite{Coish2004}.

DNSP is investigated using single self-assembled QDs in gated
structures that allow for deterministic charging of a
QD\cite{Warburton2000} with a single excess electron or hole. The
sample is grown by molecular beam epitaxy on a $(001)$
semi-insulating GaAs substrate. The InAs quantum dots are grown
25~nm above a 40~nm heavily doped n$^{+}$-GaAs layer, followed by
30 nm GaAs and 29 periods of AlAs/GaAs (2/2~nm) superlattice
barrier layer, and capped finally by 4~nm GaAs. A bias voltage is
applied between the top Schottky and back ohmic contacts to
control the charging state of the quantum dots. The density of
quantum dots is below $0.1/\mu m^{2}$, allowing the addressing of
a single quantum dot using a micro-photoluminescence ($\mu$-PL)
setup. In this letter, data based on two different QDs, labeled as
QD-A and QD-B, is analyzed.

The standard $\mu$-PL setup is based on a combination of a solid
immersion lens ($ZrO_2$, refractive index $n\approx2.2$) in
Weierstrass (or supersphere) configuration and an objective with a
numerical aperture of 0.26. A longitudinal (z-axis) magnetic field
ranging from $B_{ext}=0$ to 20~mT is produced by Helmholtz coils
positioned around the flow cryostat. The spectroscopy system
consists of a 0.75 m monochromator and a liquid-nitrogen cooled CCD
camera, providing a spectral resolution of $\sim30~\mu$eV. By using
a scanning Fabry-Perot interferometer of 15 GHz (62 $\mu$eV) free
spectral range and a finesse $>$70, a spectral resolution $<1\mu$eV
is achieved.

\begin{figure}
\includegraphics[width=\columnwidth]{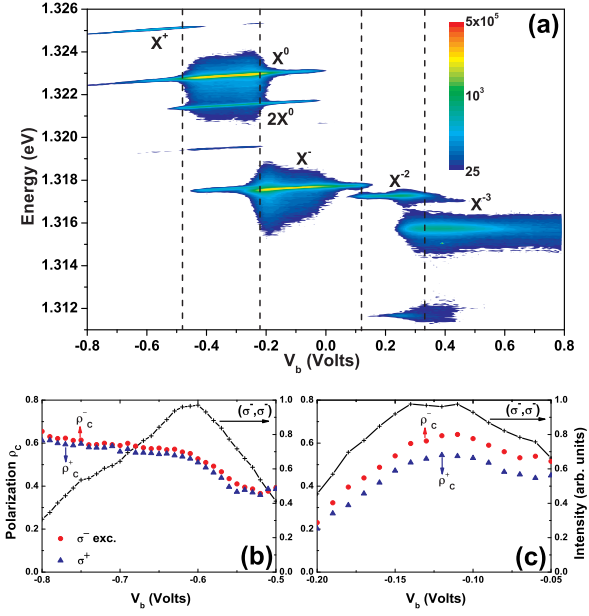}
\caption{\label{FigVbSweep} Photoluminescence from a single
charge-tunable quantum dot (QD-A). (a) Contour plot of the PL
intensity (log scale) as a function of applied bias voltage under
linear polarized excitation and detection [($\sigma^y$,$\sigma^y$)].
The excitation is at 1.35615~eV, which corresponds to the p-shell
resonance for $X^0$; $\sim$40.4~meV above $X^0$. (b) and (c) Degree
of circular polarization $\rho^\pm_c$ of PL for $X^+$ (b) and
$X^-$(c) under ($\sigma^\pm$) excitation with energy $\sim$35 and
40~meV above $X^-$ and $X^+$ lines, respectively. $\rho_c$ depends
strongly on bias voltage and weakly on pump power (not shown).}
\end{figure}

The PL polarization and spin splitting are studied by resonantly
exciting a single QD in one of its (discrete) excited (p-shell)
states under external magnetic fields ($B_{ext}$) ranging from
$B_{ext}=0$ to 20~mT, applied along the crystal growth z-axis at T =
5 Kelvin. The PL spectral lines associated with different charging
states of a single QD \cite{Warburton2000} can be identified from
the PL intensity contour plot as a function of the bias voltage and
emission energy (Fig. \ref{FigVbSweep}a). The neutral exciton $X^0$
line exhibits a fine-structure splitting of $\sim20~\mu$eV due to
the anisotropic electron-hole exchange interaction. The negatively
(positively) charged trion $X^-$ ($X^+$) emission arising from
optical excitation of a single electron (hole) charged QD is red
(blue) shifted by $\sim$5.5~meV ($\sim$3.0~meV) with respect to the
neutral exciton $X^0$ line.

The polarization for excitation and detection are denoted as
$(\sigma^\alpha,\sigma^\beta)$, where $\sigma^\alpha$ and
$\sigma^\beta$ correspond to excitation and detection,
respectively. The index $\alpha$ or $\beta$ assumes one of four
values: linear polarization along the $[110] (\sigma^y),
[1\bar{1}0] (\sigma^x)$ crystal axes or circular polarization
$\sigma^{\pm}$. The degree of circular polarization is defined as
$\rho_{c}^{\pm}\equiv (I^{\pm}-I^{\mp})/(I^{+}+I^{-})$, where
$I^\beta$ denote the intensity of PL under the
$(\sigma^{\pm},\sigma^{\beta})$ configuration. The polarization
characteristics of the system is calibrated by the Raman
scattering by the longitudinal optical phonon \cite{Yu2001} of the
GaAs substrate layer and the degree of polarization is found to be
better than $98\%$.

\begin{figure}
\includegraphics[width=\columnwidth]{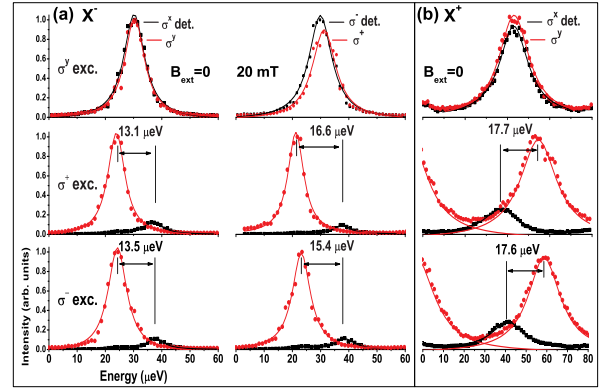}
\caption{\label{FigOS} Spin splitting induced by the Overhauser
field. (a) High-spectral-resolution $X^{-}$ PL spectra measured with
a Fabry-Perot scanning interferometer (spectral resolution
$\sim1\,\mu$eV) under $B_{ext}=0$ and 20~mT for QD-B. (b) $X^+$ PL
spectra under $B_{ext}=0$. Under circularly polarized excitation
($\sigma^{\pm}$), the detection polarization is co-circular (red
curves) or cross-circular (black) with respect to the pumping
polarization. Spin splitting of $\sim13\,\mu$eV and $\sim17\,\mu$eV
are observed under $\sigma^\pm$ excitation for $X^-$ and $X^+$,
respectively. For $B_{ext}=0$ and linearly polarized excitation
($\sigma^y$), co-linear/cross-linear polarized PL are detected. In
contrast to circular polarization, linear polarization is not
preserved: this is indicative of decoherence of hole-spin during
relaxation. The spin splitting of $X^-$ at $B_{ext}=\,20~mT$ under
linearly polarized excitation is attributed to Zeeman splitting.}
\end{figure}

Circularly polarized resonant p-shell pumping of a single electron
(hole) charged QD \cite{Bracker2005,Ware2005} generates optically
oriented trions with hole (electron) spin $J_z = 3/2$ ($S_z=-1/2$)
or $J_z = -3/2$ ($S_z=+1/2$), under $\sigma^+$ and $\sigma^-$
pumping, respectively. The intra-dot excitation ensures maximal
electron (hole) spin preservation during relaxation, which is
confirmed by the high degree of circular polarization ($\rho_c$)
of the $X^+$ ($X^-$) lines, ranging from $\sim60\%$ for QD-A (Fig.
\ref{FigVbSweep}b,c) to $\sim90\%$ for QD-B (Fig. \ref{FigOS}).
The initial state of $X^+$ trion is composed of two holes in a
singlet-state: $\sigma^+$ ($\sigma^-$) polarized PL from this
state indicates that the optically excited electron is in
$S_z=-1/2$ ($S_z=1/2$) state. For a $X^-$ trion, the initial state
is composed of an electron singlet and the PL polarization is
determined uniquely by the polarization of the hole state:
$\sigma^+$ ($\sigma^-$) polarized PL however indicates that the
electron remaining in the QD after spontaneous emission is in
$S_z=1/2$ ($S_z=-1/2$) state. In both cases polarized PL implies a
spin-polarized QD electron, which can in turn polarize nuclear
spins via hyperfine interactions\cite{Imamoglu2003}.

Figure~\ref{FigOS}a shows the spectra of $X^-$ (central energy
$\sim1.31634 $eV) for QD-B obtained using a scanning Fabry-Perot
interferometer under linearly and circularly polarized excitation.
Under linearly polarized ($\sigma^y$) laser excitation, no fine
structure splitting is observed, confirming the diminished effect
of anisotropic exchange interaction and absence of nuclear spin
polarization. Under circularly-polarized ($\sigma^{\pm}$)
excitation, spin doublets with $\sim10~\mu$eV splitting appear
even in the absence of an externally applied magnetic field. The
$X^-$ PL peaks that are co-circular with the excitation laser have
lower energies for both $\sigma^+$ and $\sigma^-$ excitation
(Fig.~\ref{FigOS}a), indicating that the direction of the
effective magnetic field responsible for the observed splitting
can be changed by switching the electron-spin polarization from
$S_z=-1/2$  to $S_z=1/2$. For $X^+$ PL (Fig. \ref{FigOS}b), this
energy sequence is reversed, indicating that the electron spin is
polarized in opposite directions in the $X^-$ and $X^+$ trions,
for a fixed circularly-polarized laser excitation \cite{Eble2005}.
Finally, measurements carried out while modulating the excitation
polarization between $\sigma^+$ and $\sigma^-$ show that the
magnitude of the spin-splitting reaches steady-state in about
$\sim1$sec: for faster modulation rates a sharp decrease in both
the spin-splitting and the PL polarization for $X^-$ is observed.
Based on these observations, we conclude that the spin-splitting
shown in Fig.~\ref{FigOS} is a clear signature of DNSP.

Coupling of a single confined electron to $N\simeq10^5$ nuclear
spins in a QD is well described by the Fermi contact Hyperfine
interaction\cite{Paget1977,Khaetskii2002}:
\begin{eqnarray}\label{Hyperfine}
\hat{H}_{hf} &=& \sum_i A_i |\psi(\textbf{R}_i)|^2 \hat{\textbf S}
\cdot \hat{\textbf I}^i \nonumber \\ &=& \sum_i \hat{B}_e^i
\hat{I}_z^i + \sum_i \frac{A_i |\psi(\textbf{R}_i)|^2}{2} [
\hat{S}_+ \hat{I}_-^i + \hat{S}_- \hat{I}_+^i] \;\;\;
\end{eqnarray}
where $|\psi({\bf R}_i)|^2$ denote the probability density of the
electron at location $\textbf{R}_i$ of the $i^{th}$ nuclear spin
and $A_i$ is the corresponding hyperfine interaction constant.
$\hat{\textbf S}$ and $\hat{\textbf I}^i$ are the electron and
nuclear spin operators, respectively. When the electron is spin
polarized via circularly polarized optical excitation under a
vanishing external magnetic field, hyperfine interactions play a
triple role: First, spin polarized confined electron leads to an
inhomogeneous Knight field $\hat{B}_e^i$ seen by each QD nucleus.
It should be emphasized that $\hat{B}_e^i$ is an operator that has
a finite mean value $\langle \hat{B}_e^i\rangle = B_e$ for a spin
polarized electron. Second, the flip-flop term $\propto \sum_i
[\hat{S}_+ \hat{I}_-^i + \hat{S}_- \hat{I}_+^i]$ in
Eq.~(\ref{Hyperfine}) enables nuclear spin pumping along the
direction determined by the electron spin, provided that electron
is continuously spin-polarized by optical excitation and that
$B_e$ is larger than local nuclear dipolar fields. Third, the
Overhauser field $B_n \propto \sum_i A_i |\psi(\textbf{R}_i)|^2
\langle \hat{I}_z^i \rangle$ induced by the polarized nuclei on
the QD electron results in a spin splitting in PL spectrum that
can be detected by high-spectral-resolution optical spectroscopy
as shown in Fig.~\ref{FigOS}.

It has been argued  that an external magnetic field exceeding the
local nuclear dipolar fields is necessary to ensure that spin
non-preserving terms in nuclear dipole-dipole interactions are
rendered ineffective in depolarizing nuclear spins\cite{Gammon2001};
this argument is correct only if the Knight field is vanishingly
small. A careful analysis shows that, if the inhomogeneous nature of
electron-nuclear coupling could be neglected, the expectation value
of the DNSP generated Overhauser field would be expressed
as\cite{Meier1984,Dyakonov1974,Paget1977,Berkovits1978}:
\begin{equation}\label{EqNuclearField}
    \textbf{B}_n= \textit{f} \, \frac{\textbf{B}^* (\textbf{B}^* \cdot \langle \textbf{S} \rangle)} {|\textbf{B}^*|^2+ \tilde{B}_L^2},
\end{equation}
where $\textbf{B}^*=B_e \hat{\textbf{z}}+\textbf{B}_{ext}$ is the
total effective magnetic field seen by the nuclei, $\langle
\textbf{S} \rangle$ is the expectation value of the electron spin,
$\tilde{B}_L$ is the effective local field characterizing nuclear
spin-spin interactions\cite{Paget1977}, and $\textit{f}$ is a
proportionality constant. In the present experiments, similar values
of the Overhauser field are observed for $B_{ext} = 0$, $20$~mT
(Fig.~\ref{FigOS}), and $200$~mT (measured using a permanent magnet;
not shown): the expectation value of the Knight field produced by a single spin-polarized electron appears to be strong enough to ensure $B_e^2 \gg
\tilde{B}_L^2$ and enables significant DNSP without an external magnetic field.

\begin{figure}
\includegraphics[width=\columnwidth]{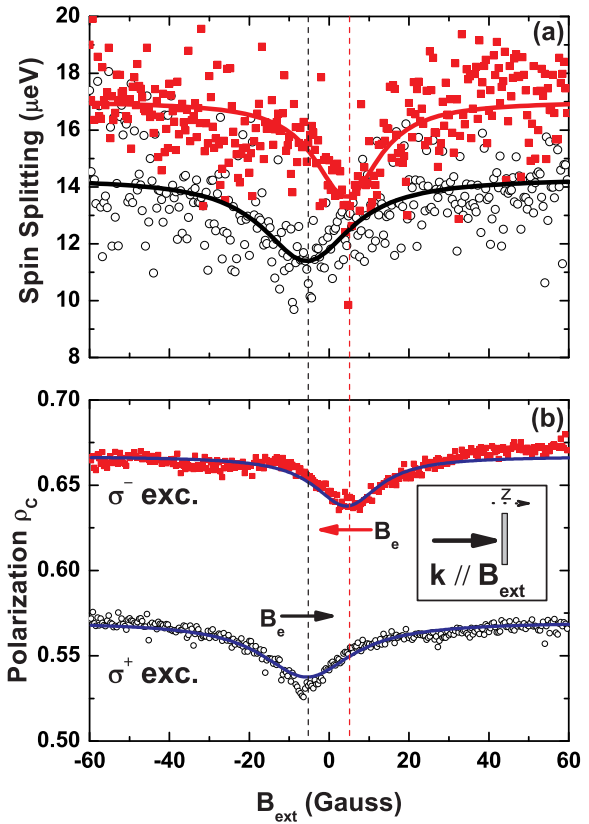}
\caption{\label{FigBSweep} Spin splitting (a) and PL polarization
(b) as a function of applied external magnetic field $B_{ext}$.
Here the spin splitting is determined by the difference of the
centroid of $X^-$ spectral lines measured by the spectrometer
(spectral resolution $\sim30\mu$eV). The solid black and red
curves in (a) are averages of the observed spin splitting.
Observation of correlated dips in spin splitting and in
polarization as a function of $B_{ext}$ suggest an average Knight
field, $B_e\simeq$6 Gauss, seen by the nuclei. Blue curves in the
polarization data are fittings by Eq~(\ref{EqPolBDep}), as
described in the text. The red and black curves in (a) are
averages of the observed spin splittings. Under $\sigma^-$
($\sigma^+$) excitation, $B_e$ is against (along) the wave-vector
$\textbf{k}$ of laser excitation and the external field
($\textbf{k}\|\textbf{B}_{ext}$).  The schematic in the inset of
(b) sketches the orientations of the laser wavevector and the
(positive) external magnetic field.}
\end{figure}

Based on Eq.~\ref{EqNuclearField}, it could be concluded that
application of an external field that cancels the Knight field (i.e.
$\textbf{B}^* = 0$) should result in the complete disappearance of
DNSP. Figure~\ref{FigBSweep}a shows the dependence of the observed
spin-splitting of $X^-$ trion under conditions where the Zeeman
splitting due to the external magnetic field ($\leq$ 50 Gauss) is
negligible: for this particular QD (A), applied gate voltage, and
the excitation intensity, a dip in spin splitting at $B_{ext} =
 -B_e\simeq +6$~Gauss is observed under $\sigma^-$ pumping.
Even at this field however, the spin-splitting  is only reduced from
$\sim16~\mu$eV to $\sim12~\mu$eV, indicating that the cancelation of
the Knight field $B_e$ by the external field is far from being
complete. The minimum in spin-splitting is observed at $B_{ext}
\simeq -6$~Gauss when the polarization of the excitation laser is
switched from $\sigma^-$ to $\sigma^+$. The observed minima, which
gives the average value of the Knight field $B_e$, ranges from
$6$~Gauss to $\sim30$~Gauss depending on the degree of PL
polarization, pumping intensity and the QD that is studied. In the
case of high $B_e$, the inhomogeneity of the Knight field, arising
from the confined electron wave-function, makes it difficult to
measure the value of $B_e$ accurately. This is due to the fact that
for any value of $B_{ext}$ most of the QD nuclei experience a strong
non-zero total magnetic field and therefore the reduction in overall
DNSP remains small.

Remarkably, a dip in the degree of PL polarization is also
observed for the same $B_{ext}$ (Fig.~\ref{FigBSweep}b): this is
at first surprising since polarization of the $X^-$ trion line is
solely determined by the hole-spin and a direct interaction
between the heavy-hole and the nuclei is unlikely to be strong
enough\cite{Gryncharova1977} to lead to the observed dependence. A
possible explanation is based on anisotropic electron-hole
exchange interaction: after the resonant excitation of the QD, the
electron excited into a p-shell state of the conduction band is
expected to tunnel out into the n-doped GaAs layer in sub-psec
timescale\cite{Smith2005}. After tunneling, the QD is neutral and
the remaining electron-hole pair is subject to anisotropic
electron-hole exchange interaction which rotates the electron-hole
spin in a correlated manner, in a timescale given by $\sim1/
\omega_{ex} \sim35$~psec (for both of the QDs studied here $\hbar
\omega_{ex} \sim20~\mu$eV). This coherent rotation is interrupted
by re-injection of another electron from the n-doped GaAs layer
into the QD s-shell to form an electron-singlet in $\tau_t\sim
5-20$~psec, as required by the charging condition. Because
tunneling is a random process with an average waiting time
$\tau_t$, the post-tunneling hole-spin state is partially
randomized and leads to a finite PL polarization. The
Overhauser-field competes with the exchange interaction; a
reduction in DNSP will therefore lead to a reduction in $\rho_c$
as depicted in Fig.~\ref{FigBSweep}b. The PL polarization in the
presence of a nuclear Overhauser shift ($\Omega_{hf} \propto B_n$)
and exchange interaction can be approximated as
\cite{Ivchenko1997}:
\begin{equation}\label{EqPolBDep}
    \rho_c=\frac{1+\Omega_{hf}^{2}\tau_{t}^2}{1+(\Omega_{hf}^{2}+\omega_{ex}^{2})\tau_{t}^2},
\end{equation}
provided other spin relaxation processes are neglected. Fitting
the polarization $\rho_c(X^-)$ with the measured spin splitting in
Fig. \ref{FigBSweep}a taken for QD-A,  $\tau_t= 30$~psec is
obtained. The magnitude of $\tau_t$ \cite{Smith2005} is consistent
with the previously reported values for $\tau_t$ \footnote{As
shown in Fig.~\ref{FigVbSweep}b, QD-A exhibits an asymmetric
polarization under $\sigma^\pm$ pumping: the origin of this
asymmetry is not clear. We observed no such asymmetry for QD-B.}.
For QD-B (Fig. \ref{FigOS}a), $\tau_t \simeq 10$~psec is obtained
for $\rho_c(X^-)\simeq90\%$ using Eq.~(\ref{EqPolBDep}). Below
saturation, a reduction in the excitation power results in a
decrease in both spin-splitting and $\rho_c$: this observation
corroborates the model described by Eq.~(\ref{EqPolBDep}).

The electron (spin) exchange with the n-doped GaAs layer also
explains how QD electron-spin pumping is achieved in a negatively
charged QD: irrespective of the pre-optical-excitation electron
state, the sequential tunneling events ensure that the QD ends up in
a trion state where the electrons form an s-shell singlet.
Preservation of hole-spin in these QDs\cite{Bulaev2005} then implies
that the post-recombination electron is always projected into the
same spin-state. Presence of a shorter barrier layer between the QD
and the n-doped GaAs could ensure that hole-spin rotation due to
exchange interaction is negligible; in this limit however,
spin co-tunneling of the remaining QD electron could reduce the
electron-spin polarization.

Some of the open questions that warrant further investigation
include the reasons for relatively low level of DNSP where only
$\sim10 \%$ of the QD nuclei appear to be polarized; differences in
the magnitude of DNSP among the four different nuclear species
present in self-assembled QDs; and the role of quadrupolar
interactions enhanced by the strain \cite{Deng2005}. By using
differential transmission measurements\cite{Hogele2005}, it should
be possible to enhance the accuracy with which the Overhauser field
can be measured by at least an order of magnitude.

We thank J. Dreiser for help with processing the sample
 and K. Karrai, O. Krebs, X. Marie, G. Salis, G. Giedke, J. Taylor,
 M. Lukin, D. Bulaev, D. Loss and A. Efros for fruitful discussions. This work is supported by NCCR-Nanoscience.

\end{document}